\documentclass[preprint,floats,aps,epsfig,nofootinbib,amssymb]{revtex4}
\usepackage{graphicx}
\usepackage{dcolumn}
\usepackage{bm}
\usepackage{mathrsfs}
\usepackage{slashed}
\usepackage{graphicx,color}
\usepackage{epsfig}
\usepackage{subfigure}
\usepackage{CJK}
\usepackage{xcolor}
\begin{document}
\begin{CJK*}{GBK}{song}
\title{Constraining  parameter space of the Little Higgs model by  data of tera-Z factory and ILC  }

\author{Xing-Dao Guo$^a$, Tai-Fu Feng$^b$, Shu-Min Zhao$^b$, Hong-Wei Ke$^c$ and Xue-Qian Li$^a$}
\affiliation{$^a$. School of Physics, Nankai University, Tianjin, China\\
$^b$. School of Physics, Hebei University, Baoding, China,\\
$^c$. School of Science, Tianjin University, Tianjin, China.
}

\date{\today}

\begin{abstract}
The Standard Model prediction on the forward-backward asymmetry
for $b\bar b$ production ($A^b_{FB}$) is well consistent with the data of LEP I at the Z-pole, but
deviates from the data at $\sqrt s=89\; 93$ GeV which are slightly away from the pole.
This deviation implies that there still is room for new physics.
We calculate the $A^b_{FB}$ at vicinity of Z-pole in the Little Higgs Model
as well as other measurable parameters such as $R_b$ and $R_c$, by which we may
constrain the parameter space of the Little Higgs Model. This can be tested
in the newly proposed tera-Z factory. With the fitted parameters we further
make predictions on $A^b_{FB}$ and $A^t_{FB}$ for $t\bar t$ production at the ILC energy.


\end{abstract}

\pacs{11.30.Er, 12.60.Jv}
\keywords{forward-backward asymmetry, one-loop diagram, Feynman amplitude}

\maketitle

\section{Introduction}

As well recognized, the hadron colliders are machines for discovery.
On other aspects, the electron-positron collider, muon-collider and
even the proposed photon collider would provide detailed information
about the discovered new physics candidates. Once some peculiar
phenomena are observed at the hadron colliders such as Tevatron or
LHC beyond the expectation of the Standard Model (SM), one is
tempted to associate them to new physics. Generally, making
confirmation is difficult, especially there are too many new physics models
available and most of them can offer a
plausible interpretation towards the new observation. One of the
reasons is that by the data obtained at hadron colliders, it is
difficult to study the details which are crucial for identifying the
new interaction and/or new particles observed in the physical
process accompanied by an enormous background. That is why people
will turn to invoke high-energy lepton colliders after successful
operation of hadron colliders, especially electron-positron
colliders which are more favorable because the technique for
building such machines are more mature.

More precisely speaking, to discover new physics, one is looking for phenomena beyond the SM
expectation through experimental measurements carried at hadron
colliders. Confirming  or at least claiming  existence of new
physics needs to measure several characteristic quantities which do
not meet the SM predictions on them at electron-positron colliders.

The forward-backward asymmetry ($A^t_{FB}$) in top-antitop production
at Tevatron is one of such measurements. In $t\bar{t}$ (here we write as $Q\bar Q$ which can be applied to the case for $b\bar b$ production)
rest frame it is defined as
\begin{equation}
A^Q_{FB}\equiv\frac{N_Q(y_Q-y_{\bar{Q}}>0)-N_Q(y_Q-y_{\bar{Q}}<0)}{N_Q(y_Q-y_{\bar{Q}}>0)+N_Q(y_Q-y_{\bar{Q}}<0)},\label{afb}
\end{equation}
where $N_Q$ is the number of heavy quarks ($t$ or $b$) and $y_Q-y_{\bar{Q}}$ is
the difference of the rapidities of the $Q$ and $\bar{Q}$,
it is Lorentz invariant and defined as
\begin{equation}
y_Q-y_{\bar{Q}}=2{\rm arctanh}(\sqrt{1-{4m_Q^2\over s}}\cos\theta),\label{rap}
\end{equation}
where $s=(p_1+p_2)^2$ with $p_1$ and $p_2$ being the momenta of
$Q$ and $\bar{Q}$.  $A_{FB}$ can be further rewritten as
\begin{equation}
A_{FB}=\frac{N_Q(\cos\theta>0)-N_Q(\cos\theta<0)}{N_Q(\cos\theta>0)+N_Q(\cos\theta<0)}\label{fb},
\end{equation}
where $\theta$ is the angle between the outgoing top quark and the
injecting proton beam. Obviously, the sign of $y_Q-y_{\bar{Q}}$ is
the same as $\cos\theta$.

The data of Tevatron at the Fermilab of
$A^t_{FB}$\cite{ta} are the following: the measurements of
the CDF and D0 Collaborations yield
$A^t_{FB}=0.158\pm0.075$\cite{Aaltonen:2011kc},
$A^t_{FB}=0.162\pm0.047$\cite{Aaltonen:2012CDFnote} and
$A^t_{FB}=0.196\pm0.065$\cite{Abazov:2011prd}, which are significantly
larger than the SM prediction $A^{SM}_{FB}
=0.089$\cite{Hollik:2011ps} for top pair production. This discrepancy would compose a hint
of existence of new physics beyond SM. Numerous models beyond SM
have been proposed to explain the deviation from the SM prediction,
and we list a few of them in our reference list as examples
\cite{Xiao:2010hm,jmc,kc,yb,elb,bb,vb,kmp,zl,bg,mig,Davoudiasl:2011tv,Atwood:2013xg}, but
definitely still many important works should also be included.

We showed in our previous work \cite{Guo1} that the deviation of the theoretical
prediction from the data can be mended in the little Higgs model (LHM). It is known that the LHM is
one of the promising models which are extensions of the SM.
Definitely, a natural tendency is to check the validity of this model at
lepton collider and furthermore to constrain the model parameters.
An ideal place for this job was the LEP experiments, especially the forward-backward
asymmetry of $b\bar b$ pair production at Z-pole is more sensitive to the model than the cross section.
One notices that at the Z-pole the SM prediction on the forward-backward
asymmetry $A^b_{FB}$ for $b\bar b$ production which is similar to the definition for the $t\bar t$ pair production, is
well consistent with the LEP data\cite{ALEPH:2005ab}, but deviates from the data at the of Z-pole vicinity energies 89 GeV and
93 GeV. Even though the absolute deviations are not extremely large, they are indeed beyond a few $\sigma$'s.
$A^b_{FB}$  was systematically calculated with
the SM in Ref.\cite{ALEPH:2005ab}, and the results show that  the gap between the theoretical
value and the experimental data is beyond $1\sim 2\sigma$ at 89 and 93 GeV. It is also noted that the errors at 89 and 93 GeV
are larger than that at Z-pole, so there 2$\sigma$
implies larger deviations. Of course the
distinction might be due to the measurement errors, but one cannot exclude
a possible contribution from new physics beyond SM (BSM), and detailed discussions will be made in the final section. Taking the difference seriously, we
hope that the $1\sim 2\sigma$ deviations can be explained by new physics BSM. Moreover, we need more accurate measurements at the vicinity of the Z-pole,
fortunately, the recently proposed tera-Z factory may play an important role to provide us more information.


On another aspect, since top quark is much heavier
than rest members of quark families, its mass could be close to the
scale of new physics BSM, so that observation on processes associated
with  top quark should be more favorable for discovering new physics.
Thus, comparing the asymmetries for top ($A^t_{FB}$) and bottom
($A^b_{FB}$) productions would be interesting. Generally speaking, the new physics would make
larger contribution to $A^t_{FB}$ than to $A^b_{FB}$, but the effect of new physics on $A^b_{FB}$ may
be sizable and can be observed at accurate measurements at ILC.

Especially, as we will show later, in the LHM there
are two extra heavy vector bosons; a heavy Z-boson and a
heavy photon. The heavy Z-boson which is much heavier than the heavy photon, mainly contributes to the asymmetry $A^t_{FB}$ rather than to
$A^b_{FB}$, whereas the heavy photon which has a mass around 83 GeV, would make substantial contributions to $A^b_{FB}$.


By a direct observation, the $A_{FB}$  is induced by  the odd power
of $\cos\theta$ in the amplitude square. Obviously, such terms imply that the
parity in the process is  violated (PV). In the SM, the parity
violation in the the process $e^+e^-\rightarrow b\bar{b}$ is due to
Z boson exchange, whose interaction with fermions has both
vector and axial vector components. For next-to-leading order (NLO),
the box diagrams also generate an asymmetry, because it is
equivalent  to a t-channel tree diagram, thus results in odd powers of
$\cos\theta$, meanwhile their interference with the photon can also enlarge
the asymmetry.


The strategy of this work is to investigate
the contributions of both SM and BSM to the asymmetries in
$e^+e^-\rightarrow b\bar b$ and $e^+e^-\rightarrow t\bar t$ with a
special BSM, i.e. the LHM which we used to explain the $A^t_{FB}$
observed at Tevatron\cite{Guo1}. The energies we set are that of the tera-Z factory
and ILC (or CLIC) respectively. Then we compare the asymmetries obtained for
$t\bar t$ and $b\bar b$ to investigate differences. Even
though we employ a special model BSM, the obtained results can make
sense about the role of BSM for the asymmetries and moreover, we can use the
data to constrain the model parameters which might be applied to other physical
processes and be further tested.


This paper is organized as follows. After this introduction, in
Section 2, we formulate the total scattering cross section,
$A^Q_{FB}$ to NLO within the frameworks of SM+LHM and as well as the measurable
$R_b$ and $R_c$. The numerical results along with all the
input parameters are presented in Section 3. The obtained results are
shown explicitly in several figures and tables. The last section is devoted to a simple
discussion and conclusion.

\section{the contributions of SM and LHM to the asymmetry up to NLO}

In this section we formulate the contributions to the
$A^Q_{FB}$ and the total cross sections for the processes of
$e^+e^-\to Q\bar{Q}$ in the framework of SM+LH  up to NLO.
The derivation in the SM at one-loop level was done a long while ago\cite{ALEPH:2005ab,cern}. We
have repeated the derivation and confirmed their numerical results for $\sqrt s$ near
the Z-pole.
Then in this work, we focus on the contribution of new physics,
concretely the LHM \cite{th}.

\subsection{SM contribution}

For completeness, we first briefly review the calculation in SM. Since we first discuss the processes at energies
near the Z-pole, $Q$ refers only $b$ quark.

The amplitude of the process $e^+e^-\rightarrow b\bar b$ at the leading order
of SM  is formulated as
\begin{equation}
\begin{array}{rl}
\mathcal{M}_1=&\bar{u}(p_4)
\gamma^{\mu}\frac{-ie}{4\sin\theta_W\cos\theta_W}(-(1-\frac{4}{3}\sin^2\theta_W)+\gamma^{5})v(p_3)\\
&\frac{-i}{s-m^2_Z}\bar{v}(p_2)
\gamma_{\mu}\frac{-ie}{4\sin\theta_W\cos\theta_W}(-(1-4\sin^2\theta_W)+\gamma^{5})u(p_1)\\
&+\bar{u}(p_4)(-ie\frac{1}{3}\gamma^{\mu})v(p_3)\frac{-i}{s}\bar{v}(p_2)(-ie\gamma_{\mu})u(p_1),
\end{array}
\end{equation}
where $\theta_W$ is the Weinberg angle, $p_1$ and $p_2$ respectively stand
for the four-momenta of the initial electron and positron, and $p_{3}$, $p_{4}$
denote the four-momenta of the final $b$ and $\bar{b}$.
In Ref.\cite{ALEPH:2005ab,cern} the contribution
has been calculated up to next-to-leading order (NLO).
The NLO contribution comes from the renormalized propagator, vertex correction,
box diagrams and QED corrections.
The first two corrections are included in
the effective vector and axial vector coupling constants of Z-fermions
($I_3^f$ is the weak isospin of the fermion f) as follows:
\begin{equation}
\begin{array}{rl}
&v_f\rightarrow (\frac{e^2}{4s_W^2c_W^2}\rho_f)^{\frac{1}{2}}(I_3^f-2Q_f \kappa_f s_W^2)\\
&a_f\rightarrow (\frac{e^2}{4s_W^2c_W^2}\rho_f)^{\frac{1}{2}}I_3^f,\\
\end{array}
\end{equation}
where $\rho_f$ and $\kappa_f$ are
\begin{equation}
\begin{array}{rl}
&\rho_f=1+\frac{3e^2}{64\pi^2 s_w^2}(\frac{m_t^2}{m_w^2}-\frac{s_w^2}{c_w^2}(\ln\frac{m_H^2}{m_w^2})-\frac{5}{6})+\Delta\rho_f\\
&\kappa_f=1+\frac{3e^2}{64\pi^2 s_w^2}(\frac{m_t^2}{m_w^2}\frac{c_w^2}{s_w^2}-\frac{10}{9}(\ln\frac{m_H^2}{m_w^2})-\frac{5}{6})+\Delta\kappa_f.\\
\end{array}
\end{equation}
For $b$ quark, $\Delta\rho_f$ and $\Delta\kappa_f$ are not negligible and can be written out as
\begin{equation}
\begin{array}{rl}
&\Delta\rho_b=-2\Delta\kappa_b\\
&\Delta\kappa_b=\frac{e^2}{64\pi^2 s_w^2}(2\frac{m_t^2}{m_w^2}).\\
\end{array}
\end{equation}
The box diagram contribution was estimated
\cite{cern} to be very small, so that can be safely neglected.
As the QED corrections, only the initial state radiation (ISR) is substantial \cite{cern} which
is expressed in term of a convolution integral for the
integrated cross section.
\begin{equation}
\begin{array}{rl}
\sigma(s)=\displaystyle\int^1_{z_0}dz H_{QED}(z,s)\sigma_{ew}(z,s),   z_0\geq\frac{4m_f^2}{s}.
\end{array}
\end{equation}
where
\begin{equation}
\begin{array}{rl}
H_{QED}(s)=&\frac{2\alpha}{\pi}(L_e-1)(1-z)^{\frac{2\alpha}{\pi}(L_e-1)-1}(1+\frac{\alpha}{\pi}(\frac{3}{2}(L_e-1)+\frac{\pi^2}{3}-\frac{1}{2}))\\
&+\frac{\alpha}{\pi}((\frac{4z}{(1+z)^2}\frac{1+z^2}{1-z}-\frac{2}{1-z})(L_e-1)-\frac{4z}{(1+z)^2}\ln\frac{4z}{(1+z)^2}),\\
&\alpha=\frac{e^2}{4\pi},    L_e=\ln\frac{s}{m_e^2}
\end{array}
\end{equation}
With these corrections, we can obtain the complete SM amplitude for the process
$e^+e^-\rightarrow b\bar b$  at NLO.

\subsection{LHM contribution}

In the LHM\cite{th}, there are four neutral bosons, two are SM photon and Z-boson
and the two extra bosons are a heavy Z-boson and a heavy photon pertaining to LHM. In our previous study\cite{Guo1},
by fitting data of Tevatron, we determined that the mass of the heavy Z-boson is much heavier than
that of the SM Z-boson, thus at the LEP energy scale
its contribution can be neglected. Meanwhile
the mass of the heavy photon is around the LEP energy scale, so would modify the values of $R_b$, $R_c$ and $A^b_{FB}$ near
the Z-pole predicted by the SM. The Lagrangian for $A_H$ coupling to fermion is written as
\begin{equation}
\mathcal{L}_{A_H}=A_H \bar{q}\gamma^{\mu}(g_v^q+g_a^q\gamma^{5})q+A_H \bar{e}\gamma^{\mu}(g_v^l+g_a^l\gamma^{5})e,
\end{equation}
and the relevant parameters are listed in table \ref{couplings} of next section.
\begin{figure}
\includegraphics[width=0.55\textwidth]{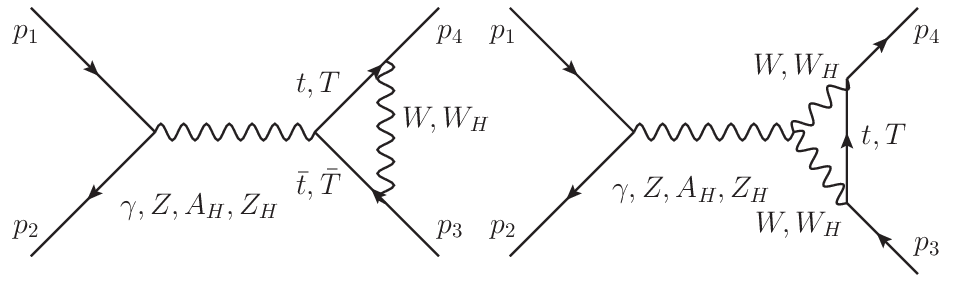}
\caption[]{The Feynman diagrams of vertex corrections for $e^+e^-\rightarrow b\bar{b}$ where $A_H$ and $Z_H$ are the
heavy photon and heavy Z-boson in LHM.}
\label{vertexlh}
\end{figure}
Similar to SM correction, the LHM vertex corrections are depicted in the Feynman diagrams of Fig.\ref{vertexlh}. To explicitly
demonstrate the procedure of  deriving the contribution, let us present the
amplitude determined by the first diagram of Fig.\ref{vertexlh} and only the
photon exchange is accounted as an example, that is:
\begin{equation}
\begin{array}{rl}
\mathcal{M}_2=&\displaystyle\int\frac{d^4k}{(2\pi)^4}\frac{-i}{k^2-m_w^2}
\bar{u}(p_4)\frac{-ie}{\sin\theta_W2\sqrt 2}V_{bt}\gamma^{\mu}(1-\gamma^{5})\\
&\frac{i(\rlap /p_4-\rlap /k+m_4)}{(p_4-k)^2-m_4^2}(-ie\frac{1}{3}\gamma_{\nu})
\frac{i(\rlap /p_3+\rlap /k+m_4)}{(p_3+k)^2-m_4^2}
\frac{-ie}{\sin\theta_W2\sqrt 2}V_{bt}\gamma^{\mu}(1-\gamma^{5})v(p_3)\\
&\bar{v}(p_2)(-ie\gamma_{\nu})\frac{-i}{s}u(p_1),\\
\end{array}
\end{equation}
where $V_{bt}$ is the CKM matrix element.
For the rest diagrams, the corresponding amplitudes can be
obtained in a similar way with different coupling
constants and masses of the intermediate fermions and bosons which are exchanged at s or t-channels.

Averaging spin projections of initial electron-positron and summing over the
spins and colors of the produced quarks, the differential cross section with respect
to the production angle $\theta$ is:
\begin{equation}\begin{array}{rl}
\frac{d{\sigma}}{d\cos\theta}&=3\times\frac{2\pi\sqrt{1-\frac{4m_Q^2}{s}}}{64\pi^2s}\frac{1}{4}
\sum|\mathcal{M}_1+\mathcal{M}_2|^2\\
&\approx 3\times\frac{2\pi\sqrt{1-\frac{4m_Q^2}{s}}}{64\pi^2s}\frac{1}{4}
(|\mathcal{M}_1|^2+2Re(\mathcal{M}_1^*\mathcal{M}_2)),
\end{array}\end{equation}
and then we integrate over the positive and
negative ranges of $\cos\theta$ respectively. The asymmetry which is expressed in terms of the Lorentz
invariant rapidity difference $y_Q-y_{\bar{Q}}$ defined in Eq.(\ref{rap}) and Eq.(\ref{afb})
is eventually derived. Moreover, we have also derived the relevant $R_b$ and $R_c$ which are
commonly defined in literature, in the SM+LHM framework. The
numerical results will be presented in next section.

\section{numerical results\label{sec3}}

Here we list all the inputs which are needed in our numerical computation. The masses of charm, bottom and top quarks
are taken as $1.27$, $4.18$ and $173.5$ GeV and the masses of light quarks
($u,\;d,\;s$) are neglected. In the center of mass frame, one has $
p_Q=p_{\bar{Q}}$ and $p_Q^2=m_Q^2$, the kinematics is determined as
\begin{equation}\begin{array}{c}
\displaystyle p_1.p_2=\frac{s}{2}, \;\;\;
\displaystyle p_3.p_4=\frac{s}{2}-m_Q^2,\\
\displaystyle p_1.p_3=p_2.p_4=\frac{s}{4}(1+\sqrt{1-\frac{4m_{Q}^2}{s}}\cos\theta),\\
\displaystyle p_1.p_4=p_2.p_3=\frac{s}{4}\left(1-\sqrt{1-\frac{4m_{Q}^2}{s}}\cos\theta\right).
\end{array}\end{equation}
For the energy of the LEP I experiment, we set
$\sqrt{s}=92.95$ GeV and $m_Z=91.2$ GeV, $m_W=80.4$ GeV
\cite{gli,data,wmy,hfr}. The electromagnetic coupling constant and
weak mixing angle are running with energy, at different energy
scales we take $\alpha_e=1/128.878$, $\sin^2\theta_W=0.2316$ for
$\sqrt{s}=91.2$ GeV; $\alpha_e=1/128.516$, $\sin^2\theta_W=0.2398$ for
$\sqrt{s}=500$ GeV; $\alpha_e=1/128.369$, $\sin^2\theta_W=0.2444$ for
$\sqrt{s}=1$ TeV \cite{pm,ac,je}. At the proposed tera-Z factory the
center-of-mass energy will be around the vicinity of $Z$ mass, so the
on-mass-shell resonance effect would be dominant and the
Breit-Winger formulation is an appropriate approach.

The coupling constants between heavy photon and various fermions are listed in
table \ref{couplings}. The mass of heavy photon is $m_{A_H}=0.08138(\frac{1}{a}+a)f$ GeV,
and $f$ is a vacuum expectation value of LHM \cite{th}.
\begin{center}
\begin{table}
\begin{tabular}[c]{|l|l|l|}
\hline
            &$g_v^q$                                      &$g_a^q$ \\\hline
$A_H\bar uu$  &$-0.0292(\frac{3}{a}-2a)$                       &$-0.0175(\frac{3}{a}-2a)$ \\\hline
$A_H\bar dd$  &$0.2742\frac{1}{a}+0.245a$                     &$0.0175(\frac{3}{a}-2a)$ \\\hline
$A_H\bar tt$  &$-0.0292(\frac{3}{a}-2a)-0.035(\frac{1}{a}+a)b$ &$-0.0175(\frac{3}{a}-2a)-0.035(\frac{1}{a}+a)b$ \\\hline
&$g_v^l$                                      &$g_a^l$ \\\hline
$A_H\bar ee$  &$0.0525(\frac{3}{a}-2a)$                       &$0.0175(\frac{3}{a}-2a)$ \\\hline
\end{tabular}
\caption{The coupling constants between heavy photon and fermion.
In the table $a=\tan\theta'$\cite{th} and $b=\frac{\lambda_1^2}{\lambda_1^2+\lambda_2^2}$
with $\lambda_1$ and $\lambda_2$ satisfy
$\frac{1}{\lambda_1^2}+\frac{1}{\lambda_2^2}\approx (\frac{v}{m_t})^2 \approx 2$\cite{th},
$v$ is the VEV of SM.\label{couplings}}
\end{table}
\end{center}

It is noted that for the heavy photon, all its couplings to fermions
uniquely depends on parameters $a$ and $b$, which are not determined
in the model, so that here we treat them as free parameters. The only way, so far before a
more fundamental principle appears, to determine them is by fitting
available experimental data.

\begin{figure}
\includegraphics[width=0.75\textwidth]{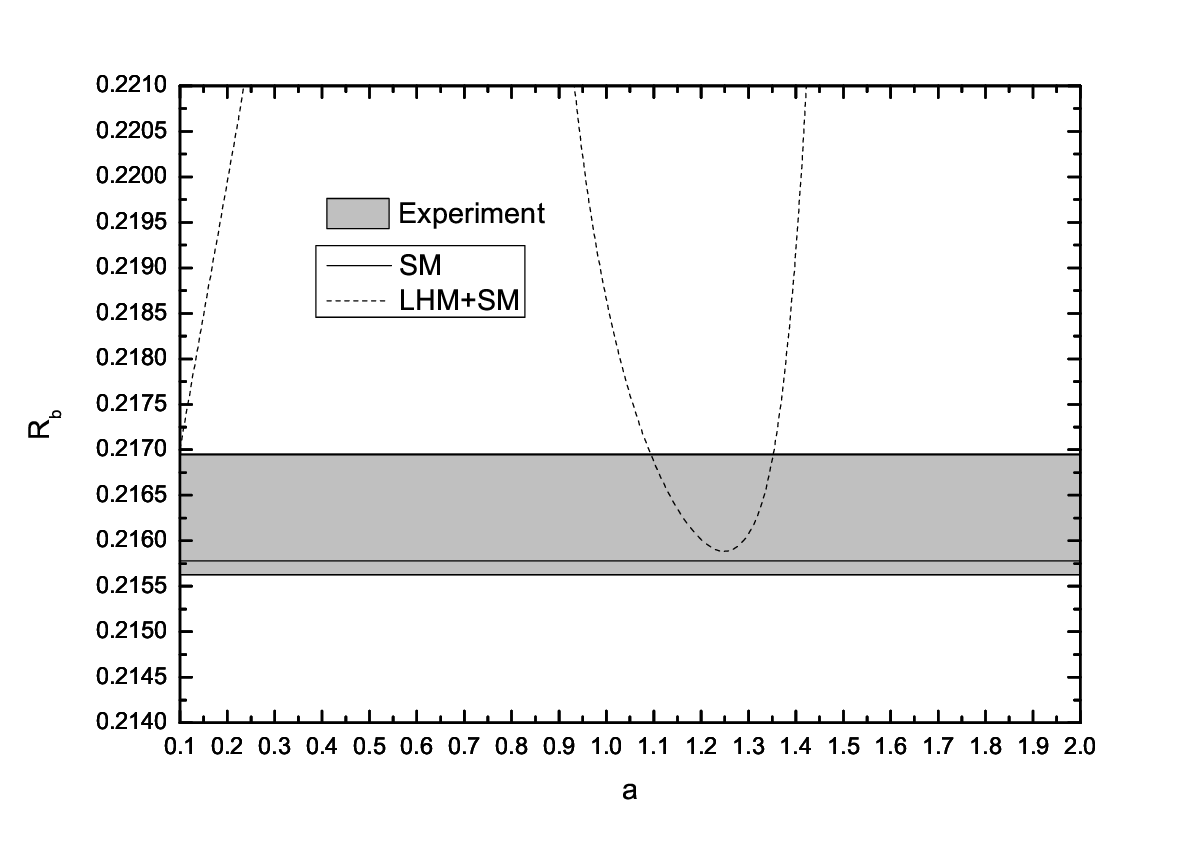}
\caption[]{Dependence of $R_b$ which is theoretically evaluated by SM-only and SM+LHM
on parameter $a$ at $\sqrt{s}=91.2$ GeV and the experimental data\cite{je}. }
\label{bratio91}
\end{figure}

\begin{figure}
\includegraphics[width=0.75\textwidth]{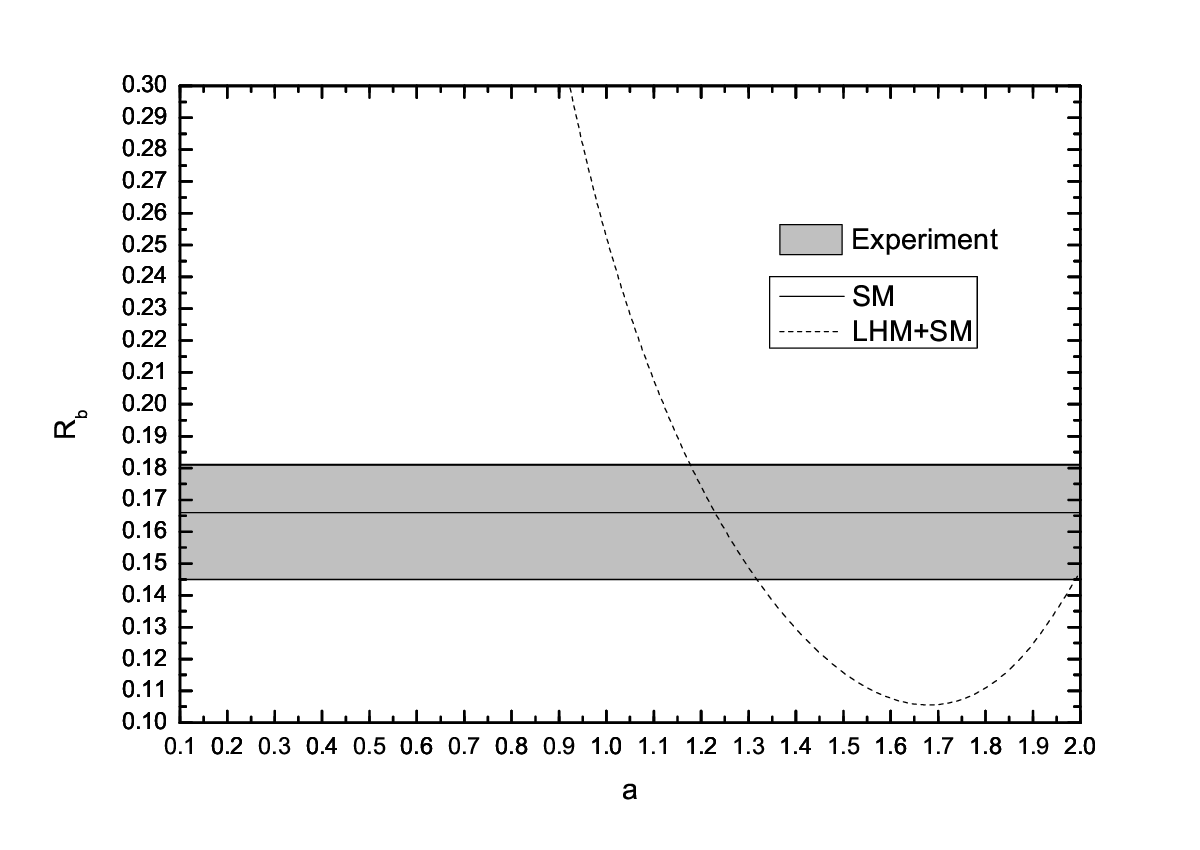}
\caption[]{Dependence of $R_b$ which is theoretically evaluated by SM-only and SM+LHM
on parameter $ a$ at
$\sqrt{s}=189$ GeV and the experimental data\cite{l3}. }
\label{bratio189}
\end{figure}

Even though the SM prediction on the asymmetry $A^b_{FB}$ is generally consistent with the LEP data,
as indicated in the introduction, there are still deviations between data and theoretical prediction as
$\sqrt s$ being away from the pole mass of Z boson. Thus we may expect that
when the contribution of LHM is included, the theoretical
prediction can be in a better agreement with experimental data, our numerical results are shown in
Fig.\ref{bratio91} and Fig.\ref{bratio189}. It is worth pointing out, for the $b\bar b$ production,
the contribution of the SM box diagrams is small, but not negligible, in comparison,
the box contributions induced by the LHM is too small to be involved, thus, we have the contributions from five sources:
the heavy photon of LHM, the box diagrams, the $\gamma$, Z boson of SM and interferences among them.

Fig.\ref{bratio91} and Fig.\ref{bratio189} show the dependence of $R_b$ which was measured at LEP I and II on
the parameter $a$ in the scenario of LHM+SM where the
ratio $R_b$ is defined as\cite{l3}:
\begin{equation}
\begin{array}{rl}
R_b=\frac{\sigma(e^+e^-\rightarrow b \bar b)}{\sigma(e^+e^-\rightarrow q \bar q)}.\label{rb}
\end{array}
\end{equation}
The results indicate that the parameter $a$ must fall into a narrow range from $1.1$ to
$1.3$ to fit the LEP I and II data.

\begin{figure}
\includegraphics[width=0.75\textwidth]{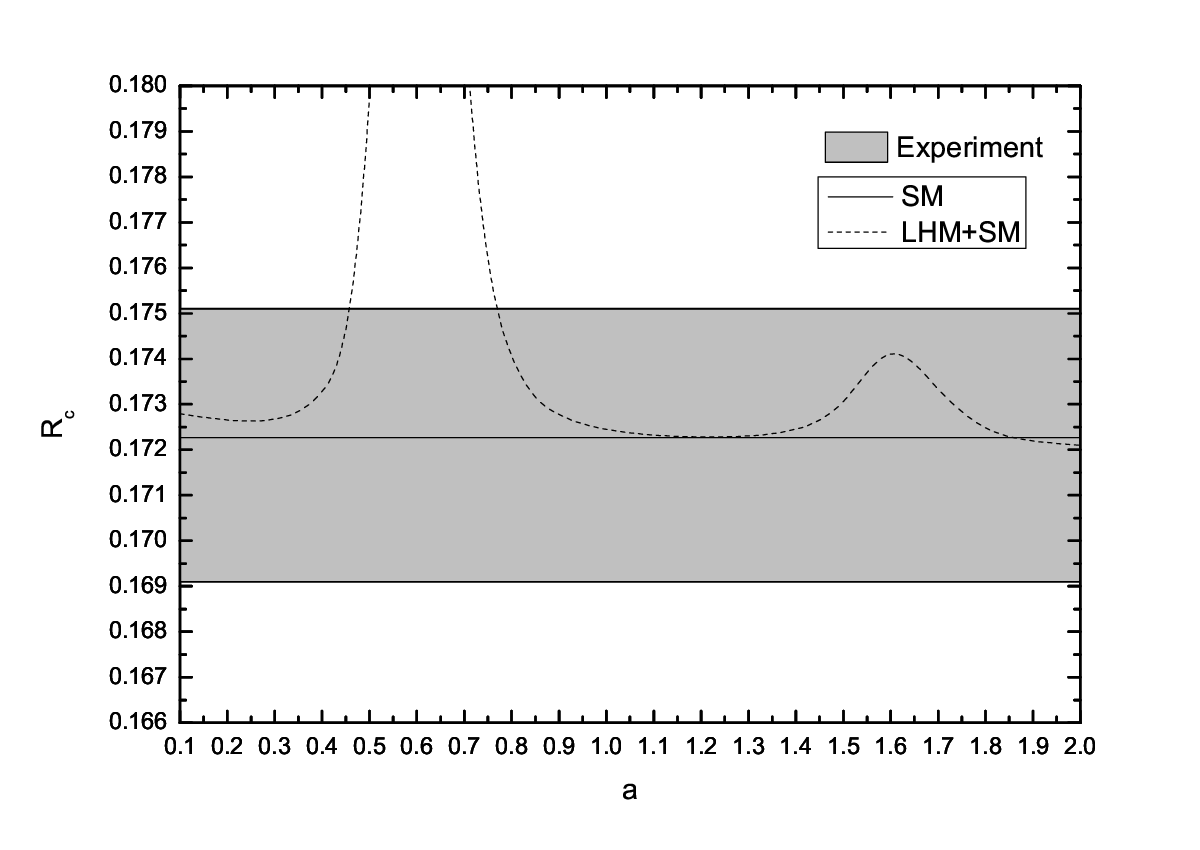}
\caption[]{The dependence of the ratio $R_c$ theoretically evaluated by SM-only and SM+LHM
on parameter $a$ at $\sqrt{s}=91.2$ GeV and the experimental data\cite{je}. }
\label{rc912}
\end{figure}

We also show $R_c$ at Z-pole versus parameter $a$
predicted by LHM+SM in Fig.\ref{rc912}.  From the results we can see that by fitting the LEP I data, there exist two windows
for the parameter $a$: $0.1\sim 0.46$ and $0.77\sim 2$.

Combining the constraints from the measured values of $R_b$and $R_c$, the parameter $a$ can be in a range of $1.1\sim 1.3$.

\begin{figure}
\setlength{\unitlength}{1mm}
\begin{center}
\begin{picture}(0,190)(20,-100)
\put(-53,0){\includegraphics{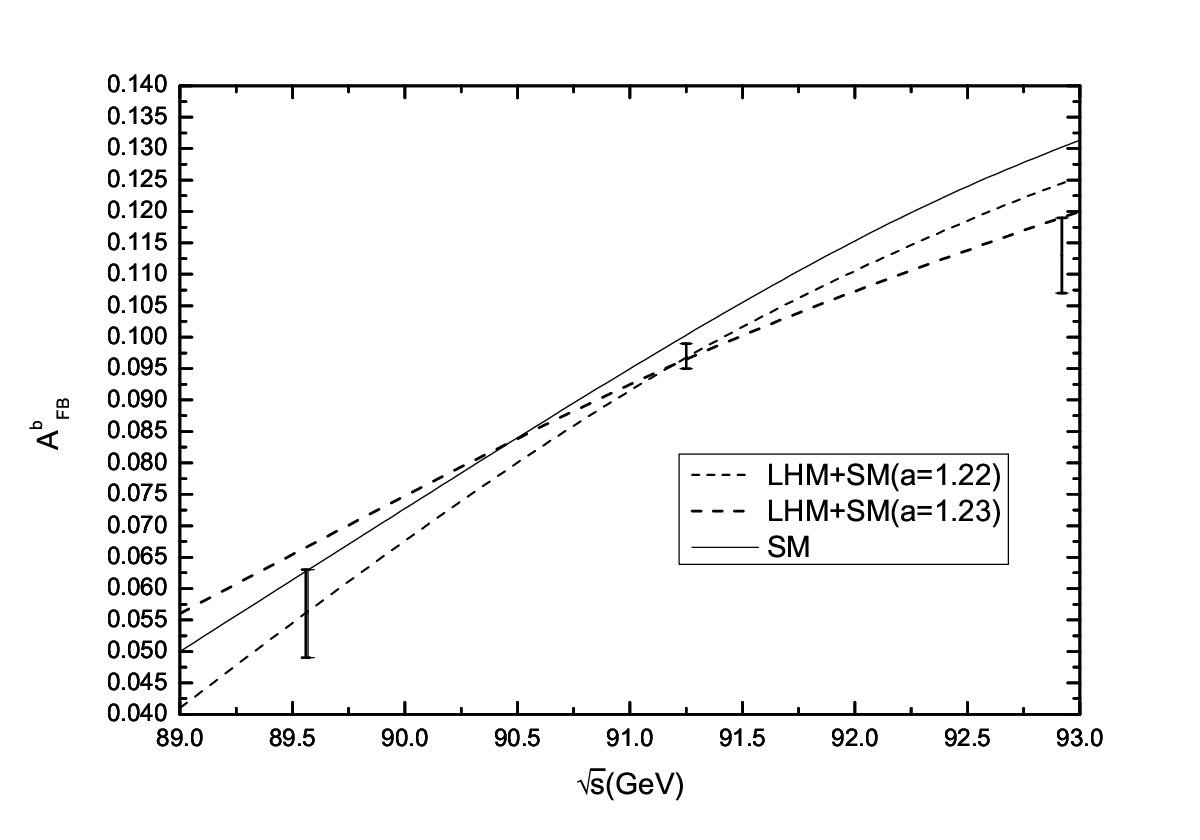}}
\end{picture}
\vspace{-10.1cm}
\caption{Dependence of $A^b_{FB}$ evaluated with SM+LHM  on the center-of-mass energy
$\sqrt{s}$ of the proposed tera-Z factory and the experimental data of LEP I \cite{ALEPH:2005ab}.
The solid line is the SM result. On the graph three experimental error bars at $89.55$ GeV $91.26$ GeV and $92.95$ GeV
respectively are explicitly shown.} \label{ahasyexp}
\end{center}
\end{figure}

In Fig.\ref{ahasyexp}, we present dependence of
$A^b_{FB}$ on $\sqrt s$ with $a$ being $1.22$ and $1.23$
respectively, where we choose   the
center of mass (CM) energy $\sqrt s$ close to the Z boson mass which is the energy range of
the proposed tera-Z factory.

\begin{figure}
\includegraphics[width=0.48\textwidth]{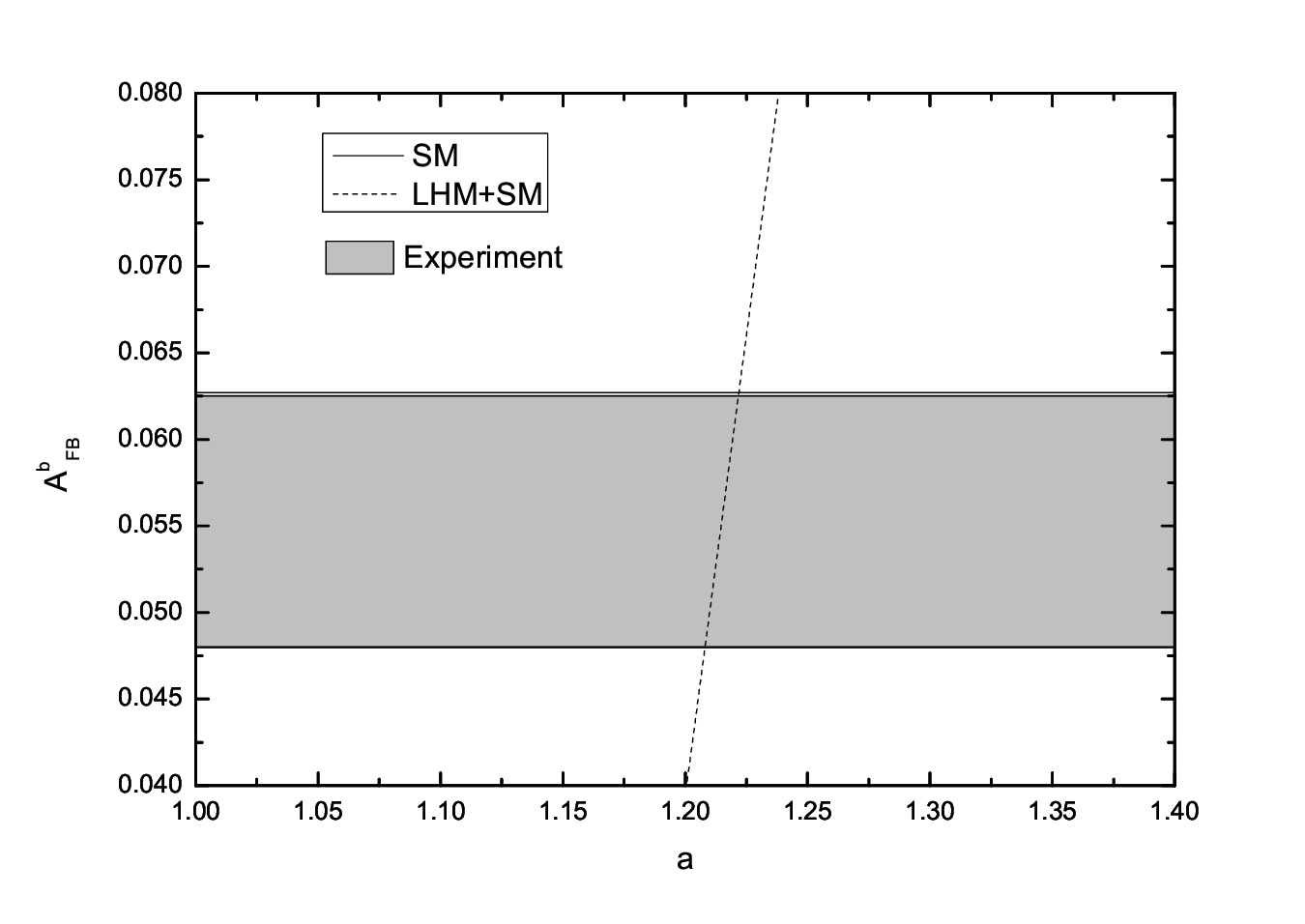}
\includegraphics[width=0.48\textwidth]{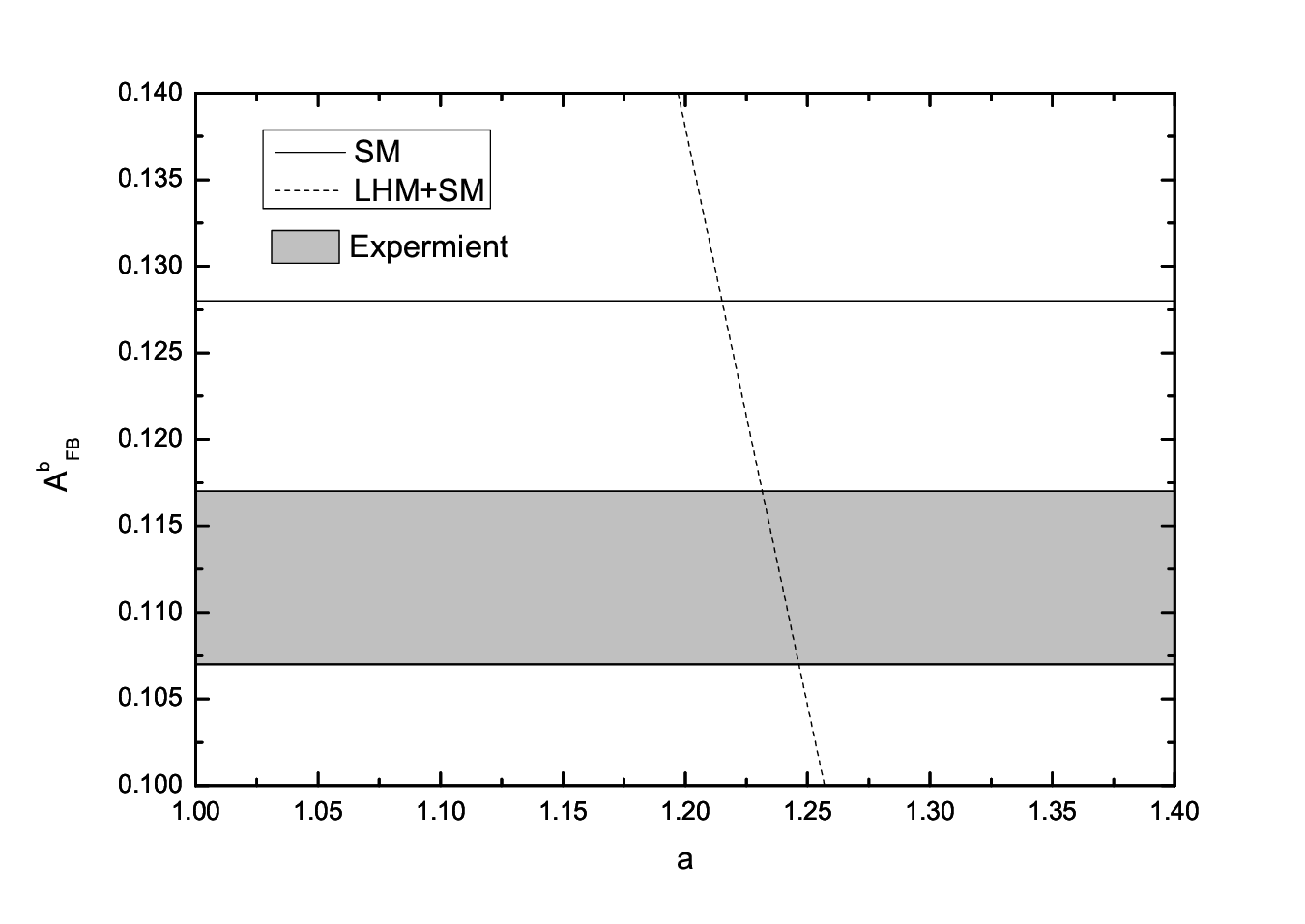}
\caption[]{The dependence of the asymmetry $A^b_{FB}$ on the parameter $a$  at
$\sqrt{s}=89.55$GeV (left) and $\sqrt{s}=92.95$GeV(right). }
\label{ahasyall}
\end{figure}

To be more explicit, let us show the the theoretical prediction on the asymmetry $A^b_{FB}$
at different CM energies
$\sqrt{s}=89.55$ GeV and $\sqrt{s}=92.95$ GeV in Fig.\ref{ahasyall} separately.
It is shown that with the LHM, agreement between the theoretical
prediction in the scenario of the SM+LHM on the asymmetry $A^b_{FB}$ and the experimental data is improved comparing with that in SM only scenario
as long as the model parameter $a$ exists in a
narrow window. However, we also observe that at $\sqrt s=M_Z$ the predicted $A^b_{FB}$ coincides well with the data, but for
center of mass energy at $\sqrt{s}=89.55$ GeV and $\sqrt{s}=92.95$ GeV, neither SM nor SM+LHM predictions can
be perfectly consistent with the data. Moreover, the theoretical estimate sensitively depends on the value of $a$.
In other words, there does not exist a common value for $a$  which can simultaneously
satisfy the measure data at $\sqrt{s}=89.55$ GeV and $\sqrt{s}=92.95$ GeV. We will discuss this point in the last section.

Now let us turn to the ILC case. For that energy range, not only the heavy photon, but also the new heavy vector boson $Z_H$ all contribute to the
asymmetry $A_{FB}^b$, moreover, since $t\bar t$ pairs can be produced, the asymmetry $A_{FB}^t$ can also be measured. $A_{FB}^t$ has been measured
at LHC \cite{Chatrchyan:2012cxa} and even though the presently available data have rather large errors, one still notices
deviations between the central values of the measured asymmetry and the SM prediction. Many authors proposed various models BSM \cite{AguilarSaavedra:2011vw,AguilarSaavedra:2011hz,Frolov:2013kpa}
to re-predict the asymmetry, more precise measurements are badly needed. In fact at the hadron colliders the signals might be contaminated by the
rather messy background, so that it is hard
to draw a definite conclusion about contributions of new physics. Therefore it is natural to expect that one can check those models at
the ILC whose low background is the most advantageous
for confirming the validity of the models and constraining the model parameter space.

\begin{figure}
\setlength{\unitlength}{1mm}
\begin{center}
\begin{picture}(0,190)(20,-100)
\put(-53,0){\includegraphics{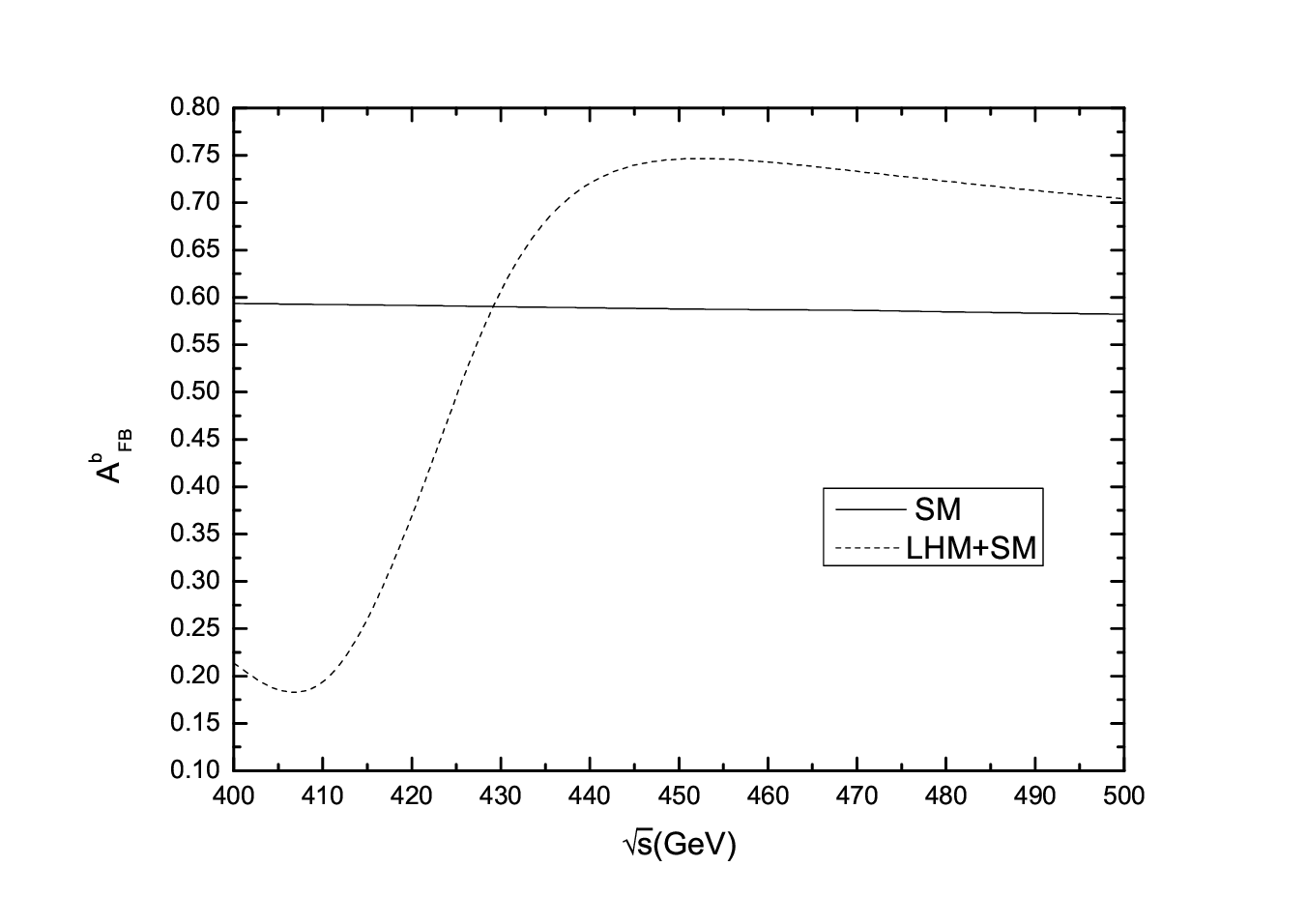}}
\end{picture}
\vspace{-10.1cm}
\caption{$A^b_{FB}$ evaluated with SM-only and LHM+SM vs the center-of-mass energy $\sqrt{s}$ of the proposed ILC.} \label{zhbbasy}
\end{center}
\end{figure}

\begin{figure}
\setlength{\unitlength}{1mm}
\begin{center}
\begin{picture}(0,190)(20,-100)
\put(-53,0){\includegraphics{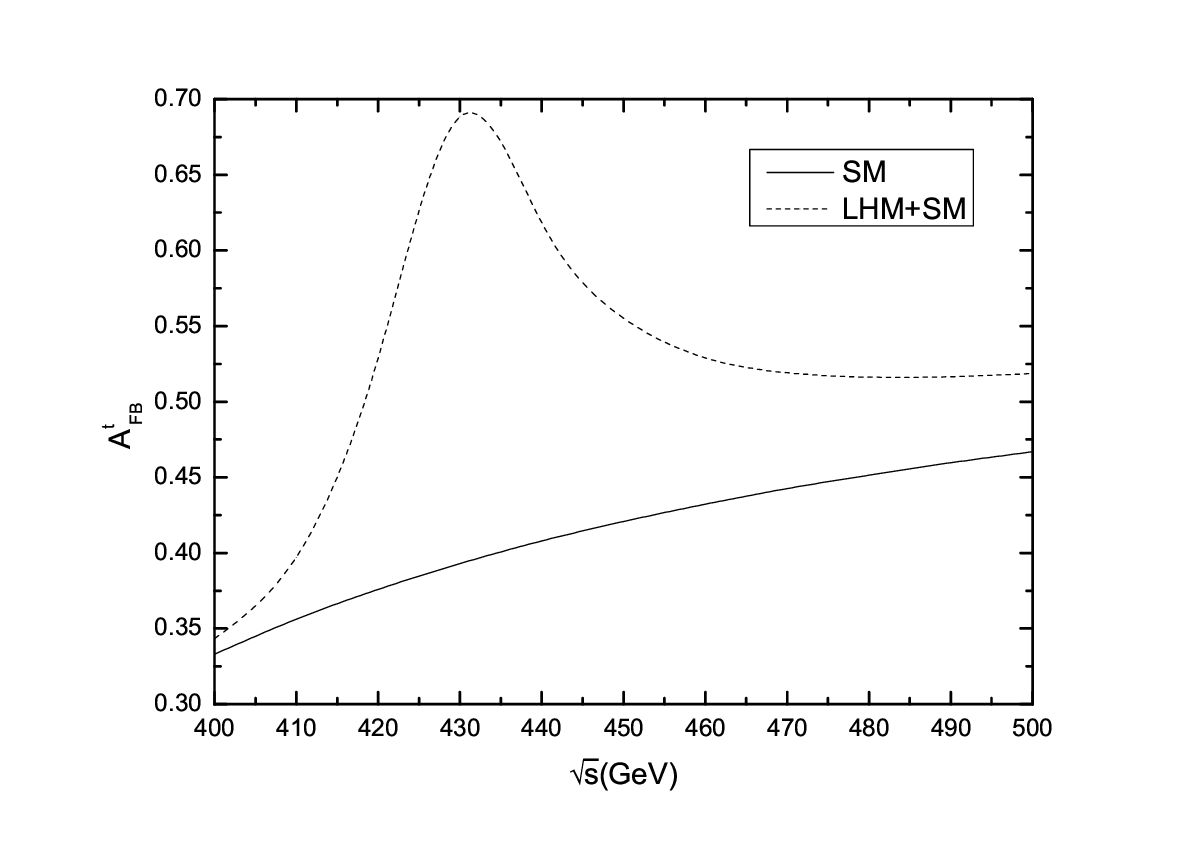}}
\end{picture}
\vspace{-10.1cm}
\caption{$A^t_{FB}$ evaluated with SM-only and LHM+SM vs the center-of-mass energy $\sqrt{s}$ of ILC.} \label{zhttasy}
\end{center}
\end{figure}

Fig.\ref{zhbbasy} and Fig.\ref{zhttasy} respectively demonstrate the dependence of
$A^b_{FB}$ and $A^t_{FB}$ on the center-of-mass energies at the proposed ILC .

Fig.\ref{zhbbasy} shows that as the mass of heavy $Z_H$ being set at 450 GeV, the $A^b_{FB}$
evaluated with LHM+SM has a minimum near $\sqrt s=410$ GeV, and a maximum at $\sqrt s=450$ GeV, this
is understood as the effects of interference between the heavy
$Z_H$, heavy photon of LHM and the SM Z boson.

We depict the dependence of the evaluated asymmetry for top pair production in Fig.\ref{zhttasy} which shows that $A^t_{FB}$ behaves quite differently
for the SM-only and LHM+SM predictions. The behavior of
$A^t_{FB}$ evaluated with LHM+SM has a bump peaked at $\sqrt s=430$ GeV,
this is also caused by an interference between
$Z_H$ and SM particles while the contribution from heavy photon can be safely ignored. As $\sqrt s$ being above $500$ GeV and
below $400$ GeV, the theoretically predicted value of $A^t_{FB}$ tends
gradually to be dominated by SM.

Table \ref{ratio} presents the ratios for the production rates of top quark $R_t$
and bottom quark $R_b$ which are defined in Eq.(\ref{rb})
at various energies, and experimental data if they are available.

\begin{center}
\begin{table}
\begin{tabular}[c]{|l|l|l|l|c|}
\hline
 &&\multicolumn{2}{c}{$R_t^{theor}$}   \vline         &$R_t^{exp}$ \\\hline
 &                               &SM            &LHM+SM           &        \\\hline
t&ILC(500GeV)         &0.165956              &0.142362             &                   \\\cline{2-5}
 &ILC(1TeV)            &0.217092               &0.205868           &                 \\\cline{1-5}
&&\multicolumn{2}{c}{$R_b^{theor}$}   \vline         &$R_b^{exp}$ \\\hline
 &ILC(500GeV)         &0.118978              &0.149088              &               \\\cline{2-5}
 &ILC(1TeV)          &0.115201               &0.124723               &               \\\cline{2-5}
b&tera-Z(92.95GeV)   &0.21406                &0.21495-0.21493         &                 \\\cline{2-5}
 &LEPI(91.2GeV)     &0.21576                   &0.21580-0.21579       &0.21629$\pm$0.00066\cite{je}\\\cline{2-5}
 &LEPII(189GeV)      &0.16035                   &0.15758-0.15757        &$0.163\pm0.013(stat)\pm0.005(syst)$\cite{l3}\\\hline
\end{tabular}
\caption{The evaluated  $R_t$
and  $R_b$. (The values predicted
by LHM correspond to $a$ varying from $1.22$ to
$1.23$)\label{ratio}}
\end{table}
\end{center}

In table \ref{asy}, we list the $A^t_{FB}$ and $A^b_{FB}$ evaluated
with SM and LHM+SM, and  available experimental data.

\begin{center}
\begin{table}[h]
\begin{tabular}{|l|l|l|l|c|}
\hline
 &               &\multicolumn{2}{c}{$A_{FB}^{t (theor)}(\%)$}\vline &$A_{FB}^{t (exp)}(\%)$                \\\cline{1-5}
 &                              &SM &LHM+SM                &                                  \\\hline
t&ILC(500GeV)                &46.68&51.86                      &                                  \\\cline{2-5}
 &ILC(1TeV)                  &56.63&54.81                      &                                  \\\cline{1-5}
 &               &\multicolumn{2}{c}{$A_{FB}^{b (theor)}(\%)$}\vline &$A_{FB}^{b (exp)}(\%)$        \\\cline{1-5}
 &ILC(500GeV)                &58.83&70.64                      &                                  \\\cline{2-5}
 &ILC(1TeV)                  &55.06&61.27                      &                                  \\\cline{2-5}
b&tera-Z(92.95GeV)   &12.84&12.41-11.71                 &                           \\\cline{2-5}
 &LEPI (91.2GeV)      &10.07&9.86-9.71                 &9.89$\pm$0.27$\pm$0.13\cite{plb577}\\\cline{2-5}
 &LEPII (189GeV)       &66.57&55.29-54,21              &$61\pm18(stat)\pm9(syst)$\cite{l3}\\\hline
\end{tabular}
\caption{Theoretical predicted $A^Q_{FB}\; (Q=t,b)$, in LHM+SM  $a$ is
in the narrow  range of $1.22\sim 1.23$. \label{asy}}
\end{table}
\end{center}

From table \ref{asy} we notice that introducing the LHM which is an extension of
SM, the discrepancy between the SM prediction on $A^b_{FB}$
and the data  at LEP energies  is alleviated. On other aspect, to fit
the data of the asymmetry, the parameter $a$ is required to fall into a rather narrow window,
it seems to be slightly fine-tuning.

Recently, successfully running of LHC enables us to analyze the contribution of LHM to $A_{FB}^t$
and $A_{FB}^b$ qualitatively. However, LHC is a proton-proton collider and for the concerned phenomena,
the main process is from the gluon fusion $gg\rightarrow q\bar q$. In that process the new gauge bosons in LHM
does not directly couple to gluons, so that besides the uncertainty brought up by the messy background,
they can only appear in loop diagrams, thus their contribution to
$A_{FB}^t$ and $A_{FB}^b$ suffers a $\frac{\alpha_s}{\pi}$ factor suppression compared to
the case at electron-positron colliders.

\section{Discussion and conclusion}

The observation of the asymmetry of top pair production $A^t_{FB}$
at Tevatron, which is obviously larger than the SM prediction,
implies possible existence of new physics BSM. Many authors
\cite{Xiao:2010hm,jmc,kc,yb,elb,bb,vb,kmp,zl,bg,mig,Davoudiasl:2011tv,Atwood:2013xg} have
tried to explain the discrepancy between theoretical predictions and
data in terms of various models BSM, and LHM is one of them.
The LHM was first proposed to cancel the quadratic divergence
induced by the SM top quark at the self-energy loop of Higgs to solve the hierarchy problem for Higgs boson.
This model besides the cancelation, has more phenomenological applications to various processes. For example,
The authors of Ref.\cite{Chen:2003fm} studied its effects on $\rho$ which
is defined as $\rho=\frac{1}{\cos^2\theta_W}\frac{M_W^2}{M_Z^2}$
and is $1$ in SM at tree level\cite{detarho}. Ref.\cite{detarho}
also presents the high order corrections to the $\rho$ parameter.

Introducing the effect of LHM, the authors of Ref.\cite{Chen:2003fm} set a constraint
on the parameter space of the LHM by fitting
the measured $\rho$ and further $\Delta\rho$ whose definition is given in Ref.\cite{detarho}.

The SM prediction of the forward backward asymmetry of bottom quark
pair production $A^b_{FB}$ can fit the LEP I data well at Z-pole, but
deviates from the data at the Z-pole vicinity energy $89.55$GeV and $92.95$GeV about $1\sim 2\sigma$.

The reason is that at Z-pole the contribution of the Z boson resonance is overwhelmingly dominant, and the other contributions from interference among SM particles and  new physics
BSM are relatively small and almost do not manifest themselves. However, when the center of mass energies
deviate from the Z-pole, the effect of those interactions becomes more significant. Then the effects of new physics would show
up at the vicinity of Z-pole mass.  Incorporating
the LHM, we find that the consistency between theoretical predictions of $A_{FB}^b$ at $89.55$ GeV and $92.95$ GeV can be improved
as the model parameter $a$ takes a value of 1.22$\sim$ 1.23. By contrast, by fitting the data of $R_b$ and $R_c$, the value of $a$ can take a
wider range of $1.1\sim 1.3$.
This fine-tuning of 1.22$\sim$ 1.23 makes us slightly uncomfortable, even though the theoretical prediction by incorporating LHM
is closer to the LEP I data.

Another observation tells us that the experimental errors (see Fig.5) at $89.55$ GeV and $92.95$ GeV are larger than that at the Z-pole, so
that we need more precise data to determine if the predictions of LHM+SM can eventually well coincide with the data, therefore
our conclusion is that more precise measurements are necessary and the proposed tera-Z factory may do a good job to provide an ideal
place for testifying the LHM.

\begin{acknowledgments}
This work is supported by the National Natural Science Foundation
of China (11275036, 11047002, 11375128), the Fund of Natural
Science Foundation of Hebei Province(A2011201118) and Natural
Science Fund of Hebei University (2011JQ05, 2007113).
\end{acknowledgments}

\end{CJK*}
\end{document}